\documentclass[twocolumn,superscriptaddress,amsmath,amssymb,aps,prl,showpacs]{revtex4-1}

\usepackage{graphicx,txfonts}
\usepackage[colorlinks, linkcolor=blue, citecolor=blue, urlcolor=blue, breaklinks=true]{hyperref}

\newcommand{\eref}[1]{Eq.~(\ref{#1})}

\newcommand{\fref}[1]{Fig.~\ref{#1}}

\newcommand{\abs}[1]{\lvert#1\rvert}

\newcommand{\ket}[1]{\lvert#1\rangle}
\DeclareMathOperator{\Tr}{Tr}

\begin{document}

\title{Dynamical symmetries and crossovers in a three-spin system with collective dissipation}

\author{S.\ Pigeon}
\affiliation{Centre for Theoretical Atomic, Molecular and Optical Physics, School of Mathematics and Physics, Queen's University Belfast, Belfast BT7\,1NN, United Kingdom}
\author{A.\ Xuereb}
\affiliation{Centre for Theoretical Atomic, Molecular and Optical Physics, School of Mathematics and Physics, Queen's University Belfast, Belfast BT7\,1NN, United Kingdom}
\affiliation{Department of Physics, University of Malta, Msida MSD2080, Malta}
\author{I.\ Lesanovsky}
\affiliation{School of Physics and Astronomy, University of Nottingham, Nottingham, NG7\,2RD, United Kingdom}
\author{J.\ P.\ Garrahan}
\affiliation{School of Physics and Astronomy, University of Nottingham, Nottingham, NG7\,2RD, United Kingdom}
\author{G.\ De Chiara}
\affiliation{Centre for Theoretical Atomic, Molecular and Optical Physics, School of Mathematics and Physics, Queen's University Belfast, Belfast BT7\,1NN, United Kingdom}
\author{M.\ Paternostro}
\affiliation{Centre for Theoretical Atomic, Molecular and Optical Physics, School of Mathematics and Physics, Queen's University Belfast, Belfast BT7\,1NN, United Kingdom}

\date{\today}

\begin{abstract}
We consider the non-equilibrium dynamics of a simple system consisting of interacting spin-$1/2$ particles subjected to
a collective damping. The model is close to situations that can be engineered in hybrid electro/opto-mechanical settings. Making use of large-deviation theory, we find a Gallavotti--Cohen symmetry in the dynamics of the system as well as evidence for the coexistence of two dynamical phases with different activity levels. We show that additional damping processes smoothen out this behavior. Our analytical results are backed up by Monte Carlo simulations that reveal the nature of the trajectories contributing to the different dynamical phases.
\end{abstract}


\maketitle

Understanding and controlling the dynamical behavior of quantum systems has seen flourishing interest~\cite{Kinoshita2006,Rabitz2009,Trabesinger2012}, propelled by theoretical and experimental progress that has made it possible to observe and manipulate such systems with unprecedented accuracy. Much attention has also been devoted recently to the notion of dynamical phase transitions in such systems, relating them to the non-analyticity of, e.g., the Loschmidt echo~\cite{Heyl2013} or the logarithm of a biased partition function in large-deviation (LD) theory~\cite{Touchette2009}, which has a natural interpretation in terms of the statistics of rare trajectories observed in experiment. The study of the dynamics of quantum systems through LD methods~\cite{Garrahan2010,Budini2011,Manzano2014} emerged recently both as an extension of the theory as applied to classical systems~\cite{Merolle2005,Lecomte2007,Baule2008,Hedges2009} and as a dynamical complement to the standard analysis of equilibrium phase transitions in many-body systems~\cite{Chandler1987}. Here, non-analyticities in the LD free-energy function of a system, extracted from the equations governing its dynamical behavior, are identified in the literature with dynamical phase transition points~\cite{Garrahan2010}. 

Following Ref.~\cite{Garrahan2010}, in this paper we are interested in studying the statistical properties of rare quantum-jump trajectories~\cite{Zoller1987} of a system that interacts with a heat bath driving the system out of equilibrium. We consider the dynamical LD properties of a simple three-spin quantum open model which departs from those recently studied in two respects: first, dissipation is due to non-classical bilinear jump operators; and second, we consider a current-like dynamical order parameter.  The two central results in the paper are: the observation of 
intermittency between dynamical phases of distinct activity, itself a consequence 
of the reducibility of the dynamics in an appropriate limit due to the collective jump operators; and the existence of a Gallavotti-Cohen symmetry in LD functions associated to the time-asymmetric order parameter, analogous to that found in driven classical systems \cite{Lebowitz1999}, which gives rise to a fluctuation theorem~\cite{Crooks1999} relating to the quantum jump rate. 
Our system therefore provides a minimal but extendible model that uncovers the effects of thermal baths and the nontrivial interplay between local and global decay channels  \cite{Olmos2014} on the non-equilibrium dynamics of a quantum system.

\begin{figure}[b]
  \includegraphics[scale=0.2]{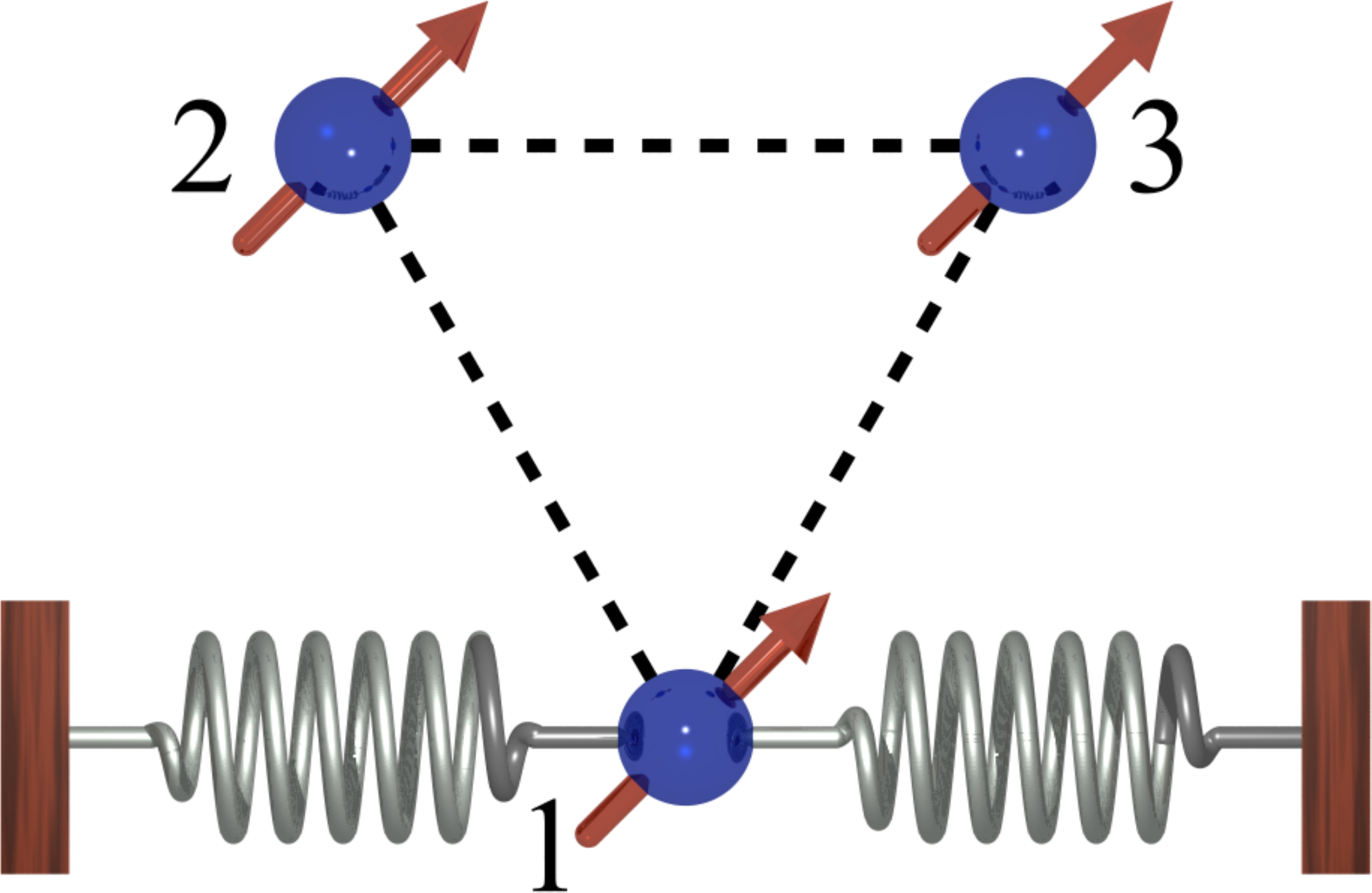}
  \caption{(Color online) The system we consider. Three spins are arranged on the vertices of an equilateral triangle. Spin $1$ is coupled to a harmonic oscillator, represented by the springs, allowing it to move in the horizontal direction. The motion of spin $1$ modulates the spin--spin interaction strengths and mediates an interaction between a collective spin degree of freedom and the mechanical bath.}
  \label{fig:System}
\end{figure}

We start with three spins-$1/2$, which we label $j=1,2,3$, placed at the vertices of an equilateral triangle (see \fref{fig:System}), interacting via an Ising-type interaction, and in a uniform magnetic field. Suppose that spin $1$ is in a harmonic trap, allowing it to move on an axis parallel to the line joining the two other spins, whereas spins $2$ and $3$ are held tightly pinned. Since the spin--spin interaction depends on the distance between each pair of spins, the motion of spin $1$ will modulate the interactions between them. By adding damping of the harmonic motion, we effectively couple the collective spin degree of freedom to this environment. Consequently, the thermal environment will act as a driving force on the spin system. More precisely, the thermal bath will drive the global degree of freedom of the spin chain. This mechanism can be generalised to other systems with similar symmetry properties with respect to the motion of a particular component. We explore the trajectories of this collective degree of freedom, finding evidence of coexisting dynamical phases and nontrivial dynamical symmetries. It is worth pointing out that individual spins can be coupled to the motion of a harmonic oscillator through the use of trapped ions~\cite{Kim2010}, by embedding solid-state qubits into mechanically-compliant structures~\cite{Yeo2014}, in nanomechanical resonator arrays~\cite{Rabl2010b}, on graphene layers~\cite{DUrso2011}, or on diamond surfaces~\cite{Cai2013}. Furthermore, this system lends itself well to being extended by adding more spins, thereby changing the symmetries of the model.

We will first derive the master equation for the spin system by adiabatically eliminating the motion of spin $1$. The resulting reduced dynamics will be investigated first by means of the quantum version of LD theory, following Ref.~\cite{Garrahan2010}. This procedure gives access to the statistics of the trajectories of the system, painting a clear picture of its dynamical behavior. Following this, we complement these analytical insights through the use of quantum jump Monte Carlo simulations \cite{Gardiner2004} that give access to a transparent physical interpretation of the processes occurring. The Hamiltonian describing the three-spin system interacting with a harmonic oscillator as described above can be written $\hat{H}=\hat{H}_\mathrm{s}+\hat{H}_\mathrm{m}+\hat{H}_{\mathrm{s}\text{--}\mathrm{m}}$ with
\begin{equation}
\hat{H}_\mathrm{s}=\alpha\sum_{i=1}^{3}\hat\sigma_\mathrm{x}^{i}+\sum_{\langle i,j\rangle}\hat\sigma_\mathrm{x}^{i}\hat\sigma_\mathrm{x}^{j}-B\sum_{i=1}^{3}\hat\sigma_\mathrm{z}^{i}\,,\\
\end{equation}
being the spin-chain Hamiltonian under uniform magnetic fields $\alpha$ and $B\ll\alpha$, $\hat{H}_\mathrm{m}=\omega_\mathrm{m}\hat{b}^\dagger\hat{b}$ the harmonic oscillator Hamiltonian, and 
\begin{equation}
\hat{H}_{\mathrm{s}\text{--}\mathrm{m}}=-g(\hat{b}^\dagger+\hat{b})\hat\sigma_\mathrm{x}^{1}(\hat\sigma_\mathrm{x}^{2}-\hat\sigma_\mathrm{x}^{3})
\end{equation}
 the interaction Hamiltonian between the spin chain and the harmonic oscillator; $\langle i,j\rangle$ denotes a sum over nearest neighbours. $\hat{H}_{\mathrm{s}\text{--}\mathrm{m}}$ follows from observing that the interaction between any pair of spins depends on the distance between them. In the geometry illustrated in \fref{fig:System}, when spin $1$ moves in a direction parallel to the line joining spins $2$ and $3$, the distance between particles $1$ and $2$ decreases (increases) by the same amount that the distance between spins $1$ and $3$ increases (decreases). Upon identifying $\hat{x}=\hat{b}^\dagger+\hat{b}$ as the dimensionless position operator for spin $1$, we arrive at the given form for the interaction Hamiltonian. Different geometries or numbers of spins can also give rise to similar Hamiltonians and effects as the ones we discuss below.
 
 We now move into the interaction picture with respect to $\hat{H}_0=\hat{H}_\mathrm{s}+\hat{H}_\mathrm{m}$, setting $\hat{b}\to\tilde{b}\,e^{-i\omega_\mathrm{m}t}$ and $\hat\sigma_\mathrm{x}^{j}  \to\tilde\sigma_{+}^{j}e^{-i Bt}+\tilde\sigma_{-}^{j}e^{i Bt}$, where $\tilde\sigma_\pm^{j}=\tilde\sigma_\mathrm{x}^{j}\pm i\tilde\sigma_\mathrm{y}^{j}$ is the spin-flip operator for the $j^{\rm th}$ spin and the tilde distinguishes interaction-picture operators from Schr\"odinger-picture ones. Assuming that $\omega_\mathrm{m}=2B$ and performing a rotating-wave approximation allows us to consider only the time-independent terms in the interaction-picture interaction Hamiltonian $\tilde{H}_{\mathrm{s}\text{--}\mathrm{m}}$, resulting in $\tilde{H}_{\mathrm{s}\text{--}\mathrm{m}}\approx-g(\tilde{b}\,\tilde\sigma_++\tilde{b}^\dagger\,\tilde\sigma_-)$, where the collective operators $\tilde\sigma_\pm=\tilde\sigma_\pm^{1}(\tilde\sigma_\pm^{2}-\tilde\sigma_\pm^{3})$. Assuming that $\abs{g}$ sets the longest time-scale of the dynamics, we can adiabatically eliminate the harmonic-oscillator degree of freedom~\cite{Gardiner2004}. To do so we follow the projection operator technique described in detail, e.g., in Ref.~\cite{Jaehne2008} and the supplemental information for Ref.~\cite{Xuereb2014}. We write a master equation, valid up to second order in $g$, that governs the evolution of the reduced interaction-picture density matrix $\tilde\rho(t)$ for the spin-only system as
\begin{equation}
 \partial_{t}\tilde\rho=\int_{0}^{\infty}\mathrm{d}\tau\,\text{Tr}_{\text{m}}\big\{\mathcal{P}_{\text{m}}\mathcal{L}_{\text{s--m}}e^{\mathcal{D}_{\text{m}}\tau}\mathcal{L}_{\text{s--m}}\mathcal{P}_{\text{m}}\tilde\rho\big\}\,,
\end{equation}
where $\mathcal{P}_{\text{m}}\tilde\rho=\tilde\rho\otimes\rho_{\text{m}}^{\text{ss}}$, with $\rho_{\text{m}}^{\text{ss}}$ the stationary solution of the harmonic oscillator, $\text{Tr}_{\text{m}}\big\{\bullet\big\}$ denotes the trace over the motional degree of freedom, $\mathcal{L}_{\text{s--m}}\bullet=-i[\tilde{H}_{\mathrm{s}\text{--}\mathrm{m}},\bullet]$, and 
\begin{align}
\mathcal{D}_{\text{m}}\bullet=&\frac\kappa2(\bar n+1)\left(2\hat{b}^\dagger \bullet \hat{b} -\{\hat{b}\hat{b}^\dagger,\bullet\}\right) \nonumber\\
& + \frac\kappa2 \bar n\left(2\hat{b} \bullet \hat{b}^\dagger -\{\hat{b}^\dagger\hat{b},\bullet\}\right)\ ,
\end{align}
 is the Lindblad-form dissipator associated with a damped harmonic oscillator connected with a damping rate $\kappa$ to a thermal bath whose average number of excitations is $\bar{n}$~\cite{Plenio1998,Gardiner2004}. The resulting master equation for the spin-system density matrix reads
\begin{equation}
\partial_t \tilde\rho =(\bar{n}+1)\Gamma\,\tilde{\mathcal{D}}_\downarrow[\tilde\rho]+\bar{n}\Gamma\,\tilde{\mathcal{D}}_\uparrow[\tilde\rho]\,, \label{eq:master}
\end{equation}
with $\Gamma=g^2/\kappa$. We also have
\begin{equation}
\tilde{\mathcal{D}}_\downarrow[\bullet]=2\tilde\sigma_{-}\bullet\tilde\sigma_{+}-\{\tilde\sigma_{+}\tilde\sigma_{-},\bullet\} \label{eq:ddown}\,,
\end{equation}
with $\tilde{\mathcal{D}}_\uparrow[\bullet]$ that is obtained from $\tilde{\mathcal{D}}_\downarrow[\bullet]$ by exchanging $\tilde\sigma_-$ and $\tilde\sigma_+$. Moving out of the interaction picture, we obtain a master equation for the reduced spin density matrix in the Schr\"odinger picture simply written as $\partial_t \rho=\mathcal{W}[\rho]$ with the corresponding super-operator $\mathcal{W}[\bullet]$ defined as
\begin{equation}
\mathcal{W}[\bullet]=-i[\hat{H}_\mathrm{s},\bullet]+\Gamma(\bar{n}+1)\,\mathcal{D}_{\downarrow}[\bullet]+\Gamma\bar{n}\,\mathcal{D}_{\uparrow}[\bullet]\label{eq:spinmaster}\,,
\end{equation}
where $\mathcal{{D}}_{\uparrow,\downarrow}:=e^{i\hat{H}_\mathrm{s}t}\tilde{\mathcal{D}}_{\uparrow,\downarrow}e^{-i \hat{H}_\mathrm{s}t}$. We similarly define $\hat\sigma_\pm$ by transforming $\tilde\sigma_\pm$ and shall drop the label $t$ when it can be understood from the context. By tracing out the motion of the damped harmonic oscillator, we have obtained an effective damping acting on a collective degree of freedom of the spin system through the jump operators $\hat\sigma_\pm$. This collective damping is the source of the interesting behavior that we shall explore in the following.

As a first step in exploring the LD behavior of our system, we associate a counting process related to the flow of excitations into or out of the system with the collective jump operators $\hat\sigma_\pm$. There are two counting processes $K_\pm$ associated with $\hat\sigma_\pm$. $K_+$ counts excitations emitted into the bath and $K_-$ excitations absorbed from it. We then define an overall counting process $K$, which counts the \emph{net} number of excitations emitted into the bath due to the collective spin flips $K:=K_+-K_-$ (in contrast to the {\em total} activity given by $K_++K_-$ \cite{Garrahan2010}). Next, we can unravel the master equation of the reduced density matrix by projecting it onto a particular number of jump events, i.e., $\partial_{t}\rho^{K}=P^{K}\mathcal{W}[\rho]$ where $P^{K}$ is a projector over trajectories with $K$ net jump events, and $p_K(t)=\Tr\{\rho^K(t)\}=\Tr\{P^K \rho(t)\}$ represents the probability to observe such a trajectory~\cite{Zoller1987}. The moment-generating function associated to this probability $p_K(t)$ can be written~\cite{Garrahan2010}:
\begin{equation}
 Z(t,s)=\sum_{K=0}^{\infty}e^{-sK}p_{K}(t)=\Tr\{\rho_{s}(t)\}\,.
\end{equation}
with $\rho_{s}(t)=\sum_{K=0}^{\infty}e^{-sK}\rho^K(t)$ the Laplace transform of the density matrix with respect to the net excitation exchanges $K$ between the system and the bath; we call $s$ the `bias parameter.' The Laplace-transformed density matrix evolves according to the modified master equation $\partial_{t}\rho_{s}=(\mathcal{W}+\mathcal{V}_{s})[\rho_{s}]$, where
\begin{equation}
\mathcal{V}_{s}[\bullet]=\Gamma\bigl[(\bar{n}+1)(e^{-s}-1)\hat\sigma_{-}\bullet\hat\sigma_{+}+\bar{n}(e^{s}-1)\hat\sigma_{+}\bullet\hat\sigma_{-}\bigr]\,.\label{eq:pretheta}
\end{equation}
In the long-time limit, LD theory applies and we can write $Z(t,s)\to e^{t\theta(s)}$, where $\theta(s)$ represents the system's dynamical free energy~\cite{Garrahan2010,Ates2012}. Consequently, we have $\theta(s)=\lim_{t\to\infty}\ln\bigl(\Tr\{\rho_{s}\}\bigr)/t$~\cite{footnote0}. $\theta(s)$ is also given by the eigenvalue of $\mathcal{W}_s:=\mathcal{W}+\mathcal{V}_s$ with the largest real part \cite{Garrahan2010} (which can be shown to be real \cite{Horssen2012}).

\begin{figure}
\begin{center}
\includegraphics[width=0.45\textwidth]{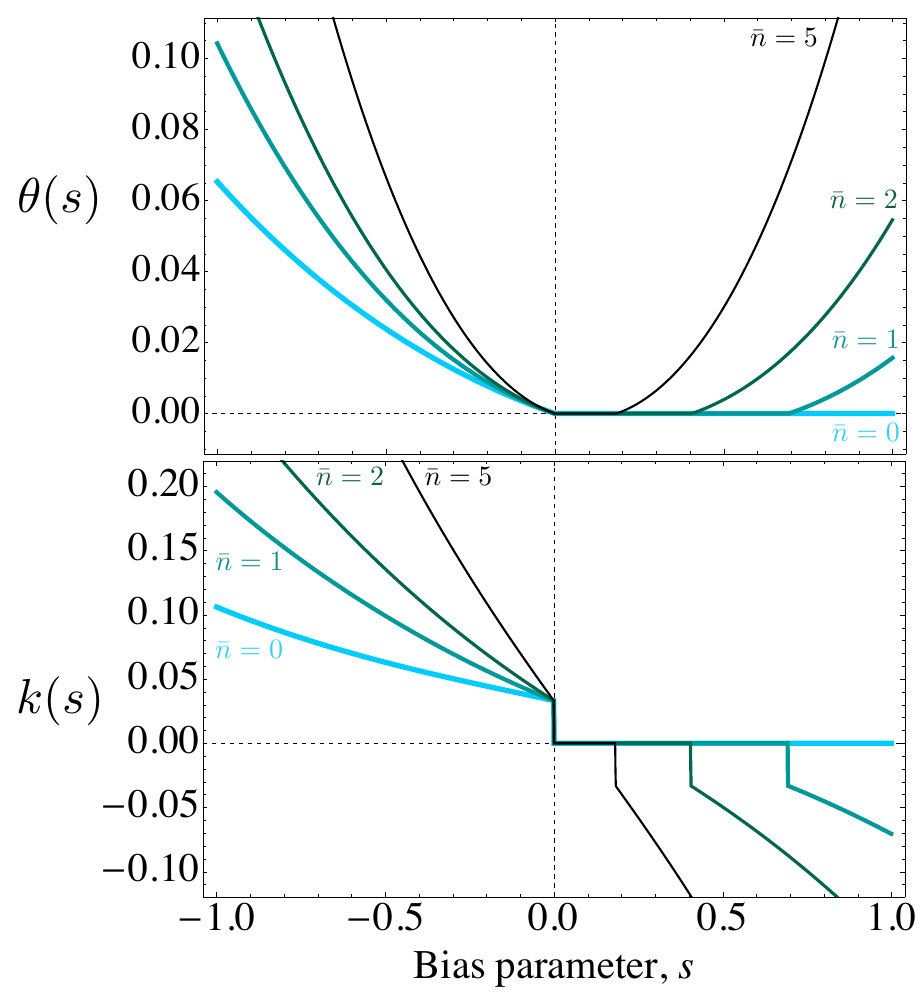}
\caption{(Color online) Illustration of the dynamical free energy $\theta(s)$ (top panel) and activity $k(s)$ (bottom). Each plot shows the quantity as a function the bias parameter $s$ for different values of the thermal population of the bath $\bar{n}$: $\bar{n}=0$, $1$, $2$ and $5$, respectively, from the thick light blue curve to the thin black one. ($\alpha=10$, $B=0.5$, and $\Gamma=0.05$.)}
\label{fig:noextradamping}
\end{center}
\end{figure}

Derivatives of this dynamical free energy with respect to $s$ can be used to obtain the activity (net count rate) $k(s)=-\partial_s \theta(s)$ of the system. This quantity is represented in \fref{fig:noextradamping} for different values of the bath population $\bar{n}$. As is clearly visible form the upper plot, for the value of the bias parameter $s=0$ we have a non-analytic point in $\theta(s)$ for any value of $\bar{n}$. The lower curves show that this point presents two distinct values of the activity $k(s)$. Unbiased dynamics takes place at $s=0$; from this we can conclude the existence of two dynamical phases~\cite{Garrahan2010,Ates2012}. These two phases have $k(0^-)>0$ and $k(0^+)=0$, i.e., one phase is active in the sense that the net rate of excitations exchanged between the bath and the system is nonzero whereas the other has an exact balance between excitations emitted and absorbed. The two terms composing the dissipative part of the master equation~\eqref{eq:spinmaster} act at different rates; we thus deduce that this balanced phase is inactive and no emission or absorption events occur. Focussing now on the active phase corresponding to $k(0^-)>0$, we also notice that the activity seems not to depend on the thermal population of the bath $\bar{n}$, since all the curves converge to the same point as $s\to0$. This effect, seen for small $\bar{n}$, stems from the weak coupling between the bath and the system ($\Gamma\ll\alpha$). 
\par
A feature of \fref{fig:noextradamping} that is not seen when considering a ``symmetric'' dynamical observable such as the total activity $K_++K_-$, as in Refs.\ \cite{Garrahan2010,Garrahan2011,Genway2012,Ates2012,Hickey2013}, is the second point of non-analyticity in $\theta(s)$ that occurs for $s>0$ when $\bar{n}>0$. Counting processes of the type we consider, unlike ``symmetric'' ones, are odd with respect to time reversal. This is related to a Gallavotti--Cohen symmetry~\cite{Lebowitz1999} due to the driven nature of the system's dynamics \cite{Speck2011}. In contrast to most studied examples of systems presenting such dynamical properties the dynamics of a \emph{global}, rather than a local, degree of freedom is considered here. Based on the detailed balance exhibited by \eref{eq:master}, we have that for $e^{s_0}=(\bar{n}+1)/\bar{n}$ we find $\theta(s)=\theta(s_0-s)$, which yields a fluctuation theorem of the form
\begin{equation}
p_K^\infty\big/p_{-K}^\infty=e^{s_0K}\xrightarrow{\bar{n}\gg1}e^{K/\bar{n}}\,,
\end{equation}
where we have defined $p_K^\infty:=\lim_{t\to\infty}p_K(t)$, that relates the infinite-time probability of observing a trajectory with a net count of $K$ to one with a net count of $-K$. This ratio approaches unity as $\bar{n}\to\infty$ in which case the rates for the collective jump process balance ($s_0\to 0$), such that $k(0^\pm)=0$~\cite{footnote}. Conversely, as $\bar{n}\to0$, the probability for observing negative $K$ goes to zero and the above ratio diverges.

\begin{figure}[t]
\begin{center}
\includegraphics[width=0.35\textwidth]{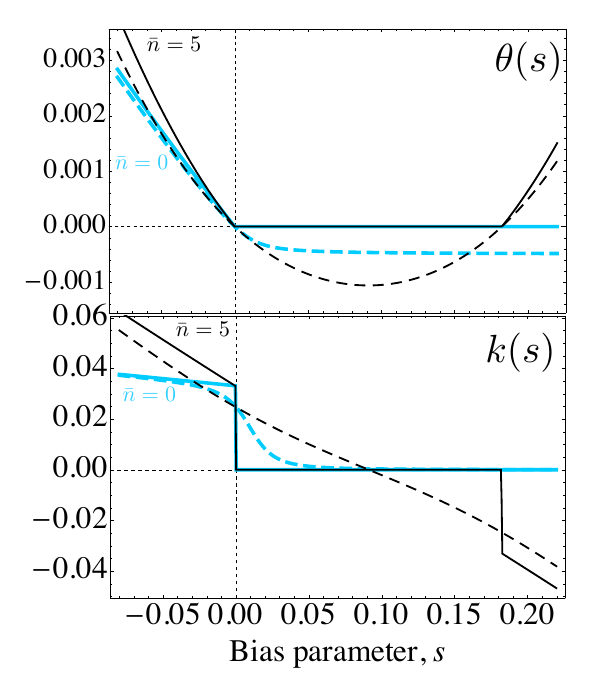}
\caption{(Color online) Illustration of the dynamical free energy $\theta(s)$ (top) and the activity $k(s)$ (bottom) without single-spin damping ($\gamma=0$; full lines) and with ($\gamma=0.01\Gamma$; dashed lines). Each plot illustrates $\bar{n}=0$ (thick light blue curve) and $\bar{n}=5$ (thin black). (Other parameters as in \fref{fig:noextradamping}.)}
\label{fig:extradamping}
\end{center}
\end{figure}

To make the model more realistic we now add independent damping channels acting on each spin and explore the consequences of these channels on the dynamical behavior of the system. For simplicity, the single-spin baths are also taken to have occupation $\bar{n}$ and be coupled with the rate $\gamma$. In~\eref{eq:spinmaster} we set $\mathcal{W}\to\mathcal{W}^\prime$:
\begin{equation}
\mathcal{W}^\prime[\bullet]:=\mathcal{W}[\bullet]+\gamma(\bar{n}+1)\sum_{i=1}^3\mathcal{D}^{i}_{\downarrow}[\bullet]+\gamma\bar{n}\sum_{i=1}^3\mathcal{D}^{i}_{\uparrow}[\bullet]\label{eq:spinmaster2}\,.
\end{equation}
$\rho_s$ evolves according to $\partial_{t}\rho_{s}=(\mathcal{W}^\prime+\mathcal{V}_{s})[\rho_{s}]$, where $\mathcal{V}_{s}[\rho_{s}]$ stays unchanged from its definition \eref{eq:pretheta} when $\gamma/3\ll\Gamma$ and neglecting correlation effects between the various damping channels. $\mathcal{D}^{i}_{\uparrow,\downarrow}$ are the spin-flip Liouvillians for spin $i$.

As \fref{fig:extradamping} illustrates, introducing extra damping at the level of the individual spins has a strong effect on the dynamical properties of the system. Concentrating on the dynamical free energy, we see that individual spin damping will smoothen out the non-analyticity at $s=0$. The same holds for the second non-analyticity at $s>0$. This smoothing effect is also visible in the activity, which becomes well-defined everywhere. Meanwhile, as it is clearly visible in \fref{fig:extradamping}, for small values of the single-spin damping and low thermal population the activity remains approximately constant. Conversely, for high thermal population and strong single-spin damping, the activity can switch sign and become negative. Physically, this corresponds to the case where the single-spin damping channel upsets the balance, making it more likely that excitations enter ($K_-$) than leave the system ($K_+$) through the collective channel, leading to a thermally driven system.

\begin{figure}[t]
\begin{center}
\includegraphics[width=0.5\textwidth]{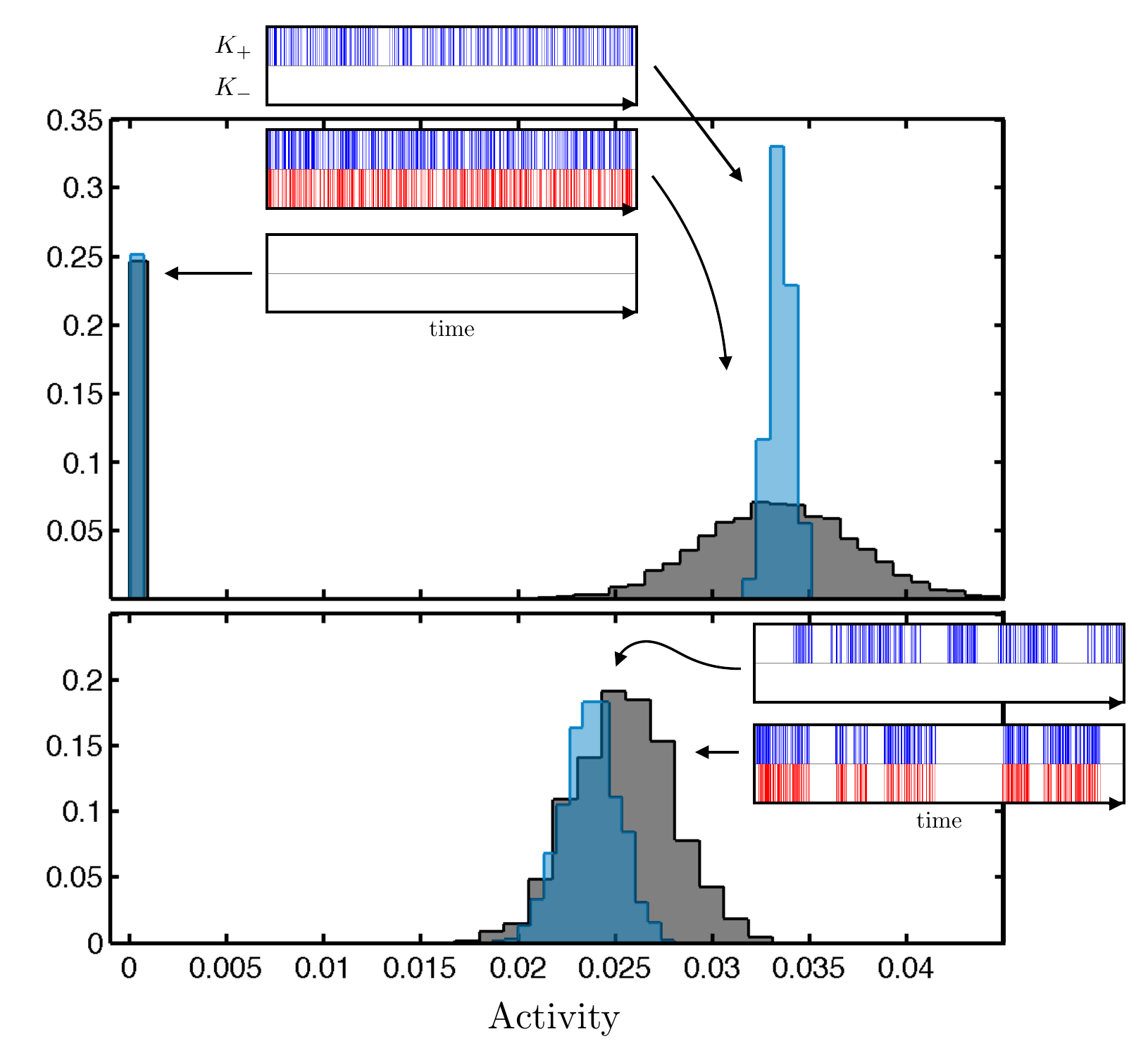}
\caption{(Color online) Probability distribution of the net activity obtained with the MCWF method, applied to an ensemble of $2000$ trajectories. The upper plot corresponds to the case without single-spin damping ($\gamma=0$), following \eref{eq:pretheta}, while the lower plot refers to $\gamma=0.01\Gamma$  [\eref{eq:spinmaster}]. In both, light blue curves show results for $\bar{n}=0$ and in dark gray for $\bar{n}=5$. The inset shows sample trajectories associated to different distributions. (Parameters as in \fref{fig:extradamping}.)}
\label{fig:wfmc}
\end{center}
\end{figure}

The LD approach to dynamical phase transitions yields a transparent physical interpretation based on the statistics of ensembles of trajectories of the system. To explore these statistics, we now conduct an analysis based on a Monte Carlo wave-function (MCWF) simulation (also known as quantum jump Monte Carlo). For classical systems, Monte Carlo simulations or other numerical methods are currently used to obtain the LD function~\cite{Lecomte2007,Hedges2009}; such methods have also occasionally been applied to quantum system~\cite{Ates2012}. The MCWF technique is a well-established method to simulate open system dynamics, like the one we are interested in, following the ideas set forth in Refs.~\cite{Gardiner2004,Dalibard1992}. Using this technique and starting off from a randomly-chosen initial condition, we simulate trajectories of \eref{eq:spinmaster2} of jump events related to the operators $\hat{\sigma}_\pm(t)$. For each trajectory generated we estimate the activity $k(0)$ by calculating the net rate of jump events. We consider a set of $2000$ trajectories, each having approximatively $10^4$ jump events, from which we obtain probability distributions for $k(0)$, as represented in \fref{fig:wfmc}. This figure shows the probability distributions for thermal populations of $\bar{n}=0$ and $5$, with (upper panel) our without (lower panel) single-spin damping. Samples of the typical trajectories obtained corresponding to different parts of the distributions are shown in inset; each vertical line corresponds to a jump event, with upper ones representing an emission from the system to the bath ($K_+$) and lower ones the opposite ($K_-$). It is clear that when $\gamma=0$ (upper panel), the probability distribution is bimodal, with one peak centred at an activity equal to $k(0^+)=0$ and the other at $k(0^-)>0$. In the former case (cf.\ inset) the corresponding trajectories have no jump events.

The second peak is centred about the same value for both values of $\bar{n}$, in agreement with what is expected from the lower panel of \fref{fig:noextradamping}, where we saw that the activity is independent of the thermal population in the low-temperature limit. Corresponding trajectories are shown in the inset, and as expected demonstrate jumps associated only with $\hat\sigma_-$ (i.e., the system losing excitations to the bath) for $\bar{n}=0$; for $\bar{n}=5$ jumps are observed in both directions. It can be see in \fref{fig:wfmc} that this second peak of the distribution broadens when the temperature increases. All this yields the interpretation that the Liouville space accessible to the system consists of two disconnected subspaces, one active and one inactive, to which the two peaks in the activity distribution are related. Based on the fraction of $K=0$ trajectories in our simulation, we can determine that $25\%$ of the Liouville space is inactive.

Consider now the case with damping on the individual spins, corresponding to the one shown in \fref{fig:wfmc} in the lower panel. By contrast to the $\gamma=0$ case, we immediately see that the distribution becomes unimodal, with the mean activity decreasing by almost $25\%$ compared to the active trajectories of the $\gamma=0$ situation. This fraction corresponds to the fraction of inactive trajectories observed when $\gamma=0$ (upper panel). The unimodal behavior and the mentioned reduction lead to the interpretation that the single-spin damping channel connects the two previously disconnected parts of the Liouville space. This interpretation is supported by the sample trajectories shown in inset of lower panel of \fref{fig:wfmc}, where one can observe the trajectory `blinking,' i.e., spontaneously switching between active and inactive behavior. In analogy with what occurs with the total activity Ref.~\cite{Ates2012}, the intermittency in trajectories is a form of ``mesoscopic'' (i.e., finite time) dynamical phase coexistence, consistent with the fact that the dynamical free-energy is analytic in this case, and the transition therefore becomes a crossover. A simple interpretation of this behaviour can be deduced by looking at the eigenspace of $\hat{H}_{\text{s}}$ and the collective spin-flip operators. We start by writing the Schr\"odinger-picture Hamiltonian $\hat{H}_{\text{s}}$ and collective operators $\hat\sigma_\pm=\hat\sigma_\pm^1(\hat\sigma_\pm^2-\hat\sigma_\pm^3)$ in the computational basis. The resulting $8\times8$ matrices are not trivial to diagonalise analytically. However, it is straightforward to find an eigenvector $\ket{\psi}$ such that $\hat\sigma_\pm\ket{\psi}=0$ and $\hat{H}_{\text{s}}\ket{\psi}=\epsilon\ket{\psi}$ for some nonzero real number $\epsilon$. The first pair of conditions render $\ket{\psi}$ a dark state of the collective spin-flip operators, which are the operators that enter the dissipative part of the reduced master equation, and the last condition assures that $\ket{\psi}$ is an eigenstate of the dynamics, i.e., that a system in $\ket{\psi}$ will remain in this inactive subspace. It can similarly be shown that no such $\ket{\psi}$ also obeys $\hat\sigma_-^1\ket{\psi}=0$, and that \emph{no state} exists such that $\hat\sigma_\pm^1\ket{\psi}=0$; in other words, any inactive state is coupled through the dynamics to the active subspace when single-spin damping is introduced. To sum up, one can find a subspace of the Liouville space that is both inactive and isolated, in the sense that $\hat{H}_{\text{s}}$ does not couple it to the rest of the space, and a subspace that is active. When $\gamma\neq0$, these two partitions are no longer isolated, and the system can switch dynamically between the active and inactive subspaces.

\begin{figure}[t]
\begin{center}
\includegraphics[width=0.49\textwidth]{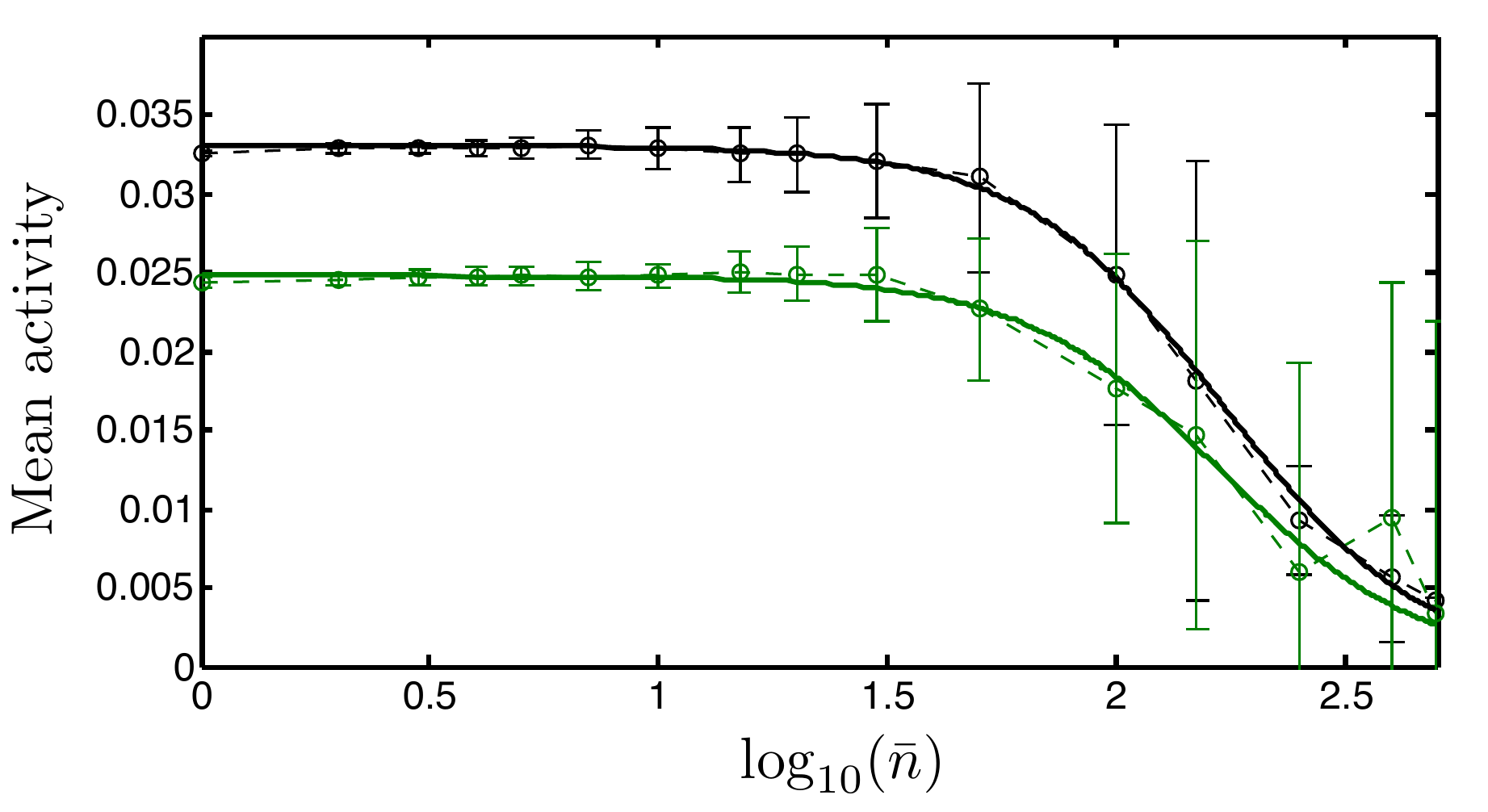}
\caption{(Color online) Activity as a function of the thermal populations $\bar{n}$ beyond the low-temperature limit. The case with (without) single-spin damping is illustrated by the lower (upper) curves in green (black). The solid curve corresponds to the large-deviation analysis and the data line to the Monte Carlo simulations. The error bars correspond to a fifth of the standard deviation. Other than $\bar{n}$, the parameters are the same as in \fref{fig:noextradamping}.}
\label{fig:wfmcvsld}
\end{center}
\end{figure}

To understand how closely our MCWF results agree with the LD analysis, we present in \fref{fig:wfmcvsld} a quantitative comparison between the two, where we plot the activity of the system at $s=0$ as a function of the thermal populations, for both low and high temperatures. This plot shows that the MCWF results (data points) agree very well with the results from the LD theory (solid curves). As visible in \fref{fig:wfmc}, increasing $\bar{n}$ tends to broaden the distribution of the activity. Consequently, the error bars shown in \fref{fig:wfmcvsld} grow quickly with $\bar{n}$. As discussed previously, we clearly see that both curves tend to zero as $\bar{n}\to\infty$, where the rates of the two counting processes $K_\pm$ balance such that the net rate of jump events is zero. Note also that the decrease in activity matches the predicted $25$\%, independently of $\bar{n}$.

We sum up by recalling our main results. We have explored a simple yet intriguing system consisting of three equidistant spins interacting pairwise, one of which moves in a harmonic trap. This motion gives rise to collective spin dynamics, which are dissipated through the mechanical decay channel. Adopting a large-deviation approach to analyse this system, we observe that its dynamics consists of two distinct dynamical phases, one active and one inactive, possessing different emission statistics. We observed that these two phases can be mixed by introducing damping on the individual spins. All our observations were confirmed through Monte Carlo simulations, which lend themselves to a natural interpretation in terms of ensembles of trajectories observed in repeated experimental runs. The system we explore is not overly complex but yields a surprisingly rich behavior. 

This work has been supported by the Royal Commission for the Exhibition of 1851, the UK EPSRC (EP/G004579/1, EP/J009776/1, EP/K029371/1 and EP/L005026/1), the John Templeton Foundation (grant ID 43467), ERC Grant ESCQUMA (Grant Agreement 335266), and the EU Collaborative Project TherMiQ (Grant Agreement 618074). Part of this work was supported by the COST Action MP1209 ``Thermodynamics in the quantum regime.''


\begin{thebibliography}{99}
\bibitem{footnote0} {The transformation from the interaction picture results in small oscillations of $Z(t,s)$ with $t$ as $t\to\infty$; we average over these oscillations when calculating $\theta(s)$ from the spectrum of the superoperator.}
\bibitem{footnote}{In this limit, therefore, barring the presence of any further non-analyticities it is a straightforward consequence of this symmetry that both $\theta(s)$ and $k(s)$ are continuous around $s=0$.}
\end{thebibliography}
\end{document}